\documentclass[journal, 12pt, onecolumn, draftclsnofoot]{IEEEtran} 

\ifCLASSINFOpdf
\else
\fi
\DeclareMathAlphabet{\mathsfbr}{OT1}{cmss}{m}{n}
\SetMathAlphabet{\mathsfbr}{bold}{OT1}{cmss}{bx}{n}
\DeclareRobustCommand{\msf}[1]{%
  \ifcat\noexpand#1\relax\msfgreek{#1}\else\mathsfbr{#1}\fi
}

\makeatletter
\newcommand{\msfgreek}[1]{\csname s\expandafter\@gobble\string#1\endcsname}
\makeatother

\DeclareFontEncoding{LGR}{}{} 
\DeclareSymbolFont{sfgreek}{LGR}{cmss}{m}{n}
\SetSymbolFont{sfgreek}{bold}{LGR}{cmss}{bx}{n}
\DeclareMathSymbol{\salpha}{\mathord}{sfgreek}{`a}
\DeclareMathSymbol{\sbeta}{\mathord}{sfgreek}{`b}
\DeclareMathSymbol{\sgamma}{\mathord}{sfgreek}{`g}
\DeclareMathSymbol{\sdelta}{\mathord}{sfgreek}{`d}
\DeclareMathSymbol{\sepsilon}{\mathord}{sfgreek}{`e}
\DeclareMathSymbol{\szeta}{\mathord}{sfgreek}{`z}
\DeclareMathSymbol{\seta}{\mathord}{sfgreek}{`h}
\DeclareMathSymbol{\stheta}{\mathord}{sfgreek}{`j}
\DeclareMathSymbol{\siota}{\mathord}{sfgreek}{`i}
\DeclareMathSymbol{\skappa}{\mathord}{sfgreek}{`k}
\DeclareMathSymbol{\slambda}{\mathord}{sfgreek}{`l}
\DeclareMathSymbol{\smu}{\mathord}{sfgreek}{`m}
\DeclareMathSymbol{\snu}{\mathord}{sfgreek}{`n}
\DeclareMathSymbol{\sxi}{\mathord}{sfgreek}{`x}
\DeclareMathSymbol{\somicron}{\mathord}{sfgreek}{`o}
\DeclareMathSymbol{\spi}{\mathord}{sfgreek}{`p}
\DeclareMathSymbol{\srho}{\mathord}{sfgreek}{`r}
\DeclareMathSymbol{\ssigma}{\mathord}{sfgreek}{`s}
\DeclareMathSymbol{\stau}{\mathord}{sfgreek}{`t}
\DeclareMathSymbol{\supsilon}{\mathord}{sfgreek}{`u}
\DeclareMathSymbol{\sphi}{\mathord}{sfgreek}{`f}
\DeclareMathSymbol{\schi}{\mathord}{sfgreek}{`q}
\DeclareMathSymbol{\spsi}{\mathord}{sfgreek}{`y}
\DeclareMathSymbol{\somega}{\mathord}{sfgreek}{`w}

\DeclareMathSymbol{\svarsigma}{\mathord}{sfgreek}{`c}

\DeclareMathSymbol{\sGamma}{\mathalpha}{sfgreek}{`G}
\DeclareMathSymbol{\sDelta}{\mathalpha}{sfgreek}{`D}
\DeclareMathSymbol{\sTheta}{\mathalpha}{sfgreek}{`J}
\DeclareMathSymbol{\sLambda}{\mathalpha}{sfgreek}{`L}
\DeclareMathSymbol{\sXi}{\mathalpha}{sfgreek}{`X}
\DeclareMathSymbol{\sPi}{\mathalpha}{sfgreek}{`P}
\DeclareMathSymbol{\sSigma}{\mathalpha}{sfgreek}{`S}
\DeclareMathSymbol{\sUpsilon}{\mathalpha}{sfgreek}{`U}
\DeclareMathSymbol{\sPhi}{\mathalpha}{sfgreek}{`F}
\DeclareMathSymbol{\sPsi}{\mathalpha}{sfgreek}{`Y}
\DeclareMathSymbol{\sOmega}{\mathalpha}{sfgreek}{`W}

\DeclareRobustCommand{\mcal}[1]{%
  \ifcat\noexpand#1\relax\mathnormal{#1}\else\cal{#1}\fi
}
\DeclareRobustCommand{\BM}[1]{%
  \ifcat\noexpand#1\relax\bm{\boldUppercaseItalicGreek{#1}}\else\bm{#1}\fi
}
\makeatletter
\newcommand{\boldUppercaseItalicGreek}[1]{\csname var\expandafter\@gobble\string#1\endcsname}
\makeatother

\newcommand{\V}[1]{\bm{#1}} 
\newcommand{\M}[1]{\BM{#1}} 
\newcommand{\Set}[1]{{\mcal{#1}}} 

\newcommand{\E}[1]{\mathbb{E}\left\{#1\right\}}
\newcommand{\T}{\mathrm{T}}

\usepackage{bm}
\usepackage{cite}
\usepackage{color}
\usepackage{psfrag}
\usepackage{acronym}
\usepackage{amsmath}
\usepackage{amssymb}
\usepackage{amsfonts}
\usepackage[colorlinks = true, allcolors = black]{hyperref}
\usepackage{latexsym}
\usepackage{mathrsfs}
\usepackage{epstopdf}
\usepackage{algorithm}
\usepackage{mathtools}
\usepackage{subfigure}
\usepackage{algorithmic}
\usepackage{relsize}
\usepackage{multirow}
\usepackage[mathscr]{euscript}

\graphicspath{{figures/}}

\newtheorem{proposition}{Proposition}
\newtheorem{remark}{Remark}
\newtheorem{theorem}{Theorem}

\DeclareMathOperator*{\argmax}{arg\,max}
\DeclareMathOperator*{\argmin}{arg\,min} 

\acrodef{iot}[IoT]{Internet of things}
\acrodef{iid}[i.i.d.]{independent and identically distributed}
\acrodef{cpu}[CPU]{central processing unit}
\acrodef{csi}[CSI]{channel state information}
\acrodef{nfv}[NFV]{network function virtualization}
\acrodef{sdn}[SDN]{software defined network}
\acrodef{ldp}[LDP]{Lyapunov Drift-Plus-Penalty}
\acrodef{mdp}[MDP]{Markov decision process}
\acrodef{mec}[MEC]{mobile edge computing}
\acrodef{agi}[AgI]{augmented information}
\acrodef{jsdm}[JSDM]{joint spatial division and multiplexing}
\acrodef{mecnc}[MECNC]{MEC Network Control}
\acrodef{cdf}[CDF]{cumulative density function}
\acrodef{pdf}[pdf]{probability density function}
\acrodef{wrt}[w.r.t.]{with respect to}
\acrodef{ue}[UE]{user equipment}

\begin{document}

\title{Mobile Edge Computing Network Control: Tradeoff Between Delay and Cost}

\author{
    Yang~Cai,~\IEEEmembership{Student~Member,~IEEE},
    Jaime~Llorca,~\IEEEmembership{Member,~IEEE},
    Antonia~M.~Tulino,~\IEEEmembership{Fellow,~IEEE}, and
    Andreas~F.~Molisch,~\IEEEmembership{Fellow,~IEEE}%
    \thanks{This paper was presented at IEEE GlobeCOM 2020 \cite{cai2020mec}. DOI: \url{10.1109/GLOBECOM42002.2020.9322331}}%
    \thanks{Y. Cai and A. F. Molisch are with the Department of Electrical Engineering, University of Southern California, Los Angeles, CA 90089, USA (e-mail: yangcai@usc.edu; molisch@usc.edu).}%
    \thanks{J. Llorca is with New York University, NY 10012, USA (e-mail: jllorca@nyu.edu).}%
    \thanks{A. M. Tulino is with New York University, NY 10012, USA, and also with the University\`{a} degli Studi di Napoli Federico II, Naples 80138, Italy (e-mail: atulino@nyu.edu; antoniamaria.tulino@unina.it).}%
    \thanks{\copyright~2020 IEEE. Personal use of this material is permitted. Permission from IEEE must be obtained for all other uses, including reprinting/republishing this material for advertising or promotional purposes, collecting new collected works for resale or redistribution to servers or lists, or reuse of any copyrighted component of this work in other works.}%
}

\maketitle

\begin{abstract}
As \ac{mec} finds widespread use for relieving the computational burden of compute- and interaction-intensive applications on end user devices, 
understanding the resulting delay and cost performance is drawing significant attention. 
While most existing works focus on single-task offloading in single-hop \ac{mec} networks, next generation applications (e.g., industrial automation, augmented/virtual reality) require advance models and algorithms for dynamic configuration of multi-task services over multi-hop MEC networks.
In this work, we leverage recent advances in dynamic cloud network control to provide a comprehensive study of the performance of multi-hop \ac{mec} networks,
addressing the key problems of multi-task offloading, timely packet scheduling, and joint computation and communication resource allocation. 
We present a fully distributed algorithm based on Lyapunov control theory that achieves throughput optimal performance with delay and cost guarantees. Simulation results validate our theoretical analysis and provide insightful guidelines on overall system design and the interplay between communication and computation resources.
\end{abstract}

\IEEEpeerreviewmaketitle

\section{Introduction}

Resource- and interaction-intensive applications such as augmented reality, real-time computer vision will increasingly dominate our daily lives \cite{cai2022metaverse}.
Due to the limited computation capabilities and restricted energy supply of end \acp{ue}, many resource-demanding tasks that cannot be executed locally end up being offloaded to centralized cloud data centers.
However, the additional delays incurred in routing data streams from \acp{ue} to distant clouds significantly degrade the performance of real-time interactive applications. To address this challenge, \acf{mec} emerges as an attractive alternative by bringing computing resources to edge servers deployed close to end users (e.g., at base stations), which allows easier access from the \acp{ue}, as well as reducing the computation expenditure.

\begin{figure}
  \centering
  \includegraphics[width = .8 \textwidth]{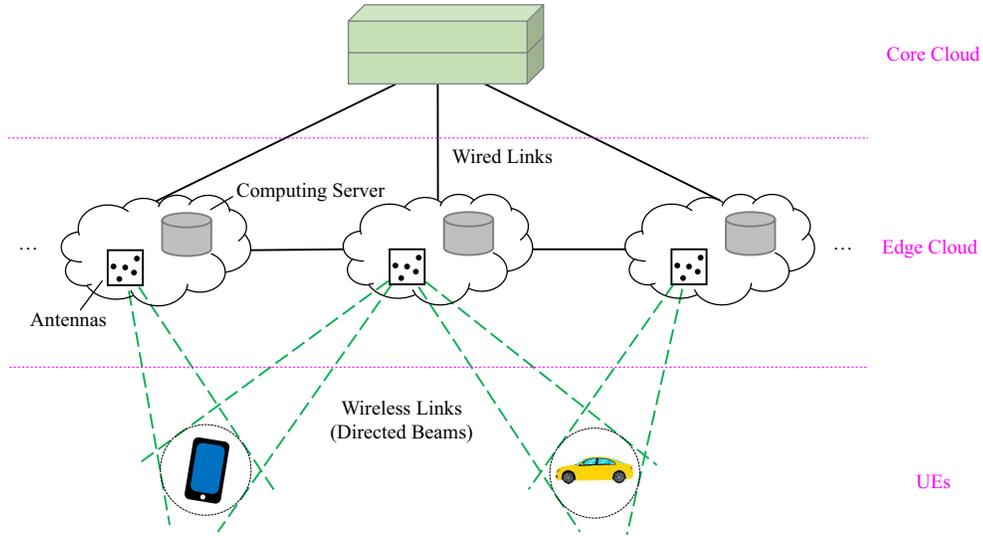}
  \caption{An illustrative multi-hop \ac{mec} network consisting of \acp{ue}, edge cloud servers, and the core cloud. Edge servers communicate with \acp{ue} via wireless links, and among each other and the core cloud via wired connections.} 
  \label{fig:mec}
\end{figure}

\begin{figure}
  \centering
  \includegraphics[width = .8 \textwidth]{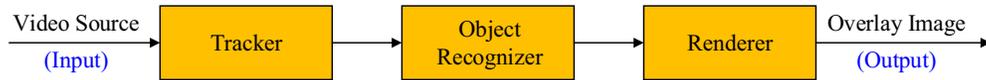}
  \caption{A function chain representation of an augmented reality service, composed of Tracker, Object Recognizer, and Renderer functions \cite{mach17mecsurvey}.}
  \label{fig:service}
\end{figure}

Delay and cost are hence two crucial criteria when evaluating the performance of \ac{mec} networks.
Offloading intensive tasks to the cloud reduces overall resource cost (e.g., energy consumption) by taking advantage of more efficient and less energy-constrained cloud servers, at the expense of increasing the delay. 
In order to optimize such cost-delay tradeoff, \ac{mec} operators have to make critical decisions including:
\begin{itemize}
\item {\em Task offloading:}  decide whether a task should be processed at the \ac{ue} (locally) or at the edge cloud, and in which edge server; 
\item {\em Packet scheduling:} decide how to route incoming packets to the appropriate servers assigned with the execution of the corresponding tasks;
\item {\em Resource allocation:}  determine the amount of computation resources to allocate for the execution of tasks at edge servers and the amount of communication resources (e.g., transmission power) to allocate for the transmission of data streams through the MEC network. 
\end{itemize}

Some of these problems have been addressed in existing literature. 
We refer the readers to \cite{DinLeeNiyWan:J13, mach17mecsurvey} and the references therein for an overview of recent MEC studies. 
For example, the task offloading problem for minimizing average delay under \ac{ue} battery constraints is addressed in \cite{CheHao:J18}; 
the dual problem of minimizing energy consumption subject to worst-case delay constraint, is studied in \cite{cai2021delay,LyuTiaNiZhaZhaLiu:J18}; \cite{TraPom:J19} investigates the same problem under an objective function that trades off the two criteria.
However, most existing works on \ac{mec} focus on simplified versions of a subset of the problems listed above, lacking a framework that jointly optimizes these aspects.

Recent advances in network virtualization and programmability have enabled even more complex services including multiple tasks (see Fig. \ref{fig:service}), which allow the individual tasks to be completed at multiple cloud locations.
A number of recent works have addressed the placement of service function chains over the cloud network \cite{BarLloTulRam:C15,BarChoAhmBou:C15,BhaJaiSamErb:J16}.
However, they generally leave out aspects such as uncertain channel conditions, time-varying service demands, and delay optimization, which are critical to \ac{mec} networks.

Driven by the proliferation of these multi-task services, in this work we focus on the design of dynamic control policy for multi-hop MEC networks hosting multi-task services (see Fig. \ref{fig:mec}). 
To achieve this goal, we leverage recent developments of applying Lyapunov optimization theory in distributed computing network management \cite{cai2022multicast_arxiv,cai2021multicast,FenLloTulMol:J18a,FenLloTulMol:J18b}.
Our contributions can be summarized as follows:

\begin{itemize}
  \item We develop an efficient algorithm for dynamic \ac{mec} network control, which jointly solves the problems of multi-task offloading, packet scheduling, and computation/communication resource allocation;
  \item We prove that the proposed algorithm is throughput-optimal while striking graceful cost-delay tradeoff,
and validate our claims via numerical simulations.
\end{itemize}

\section{System Model}

Consider a \ac{mec} network, as is shown in Fig. \ref{fig:mec}. Let $\Set{V}_\text{a}$ and $\Set{V}_\text{b}$ be the sets of {\em \acp{ue}} and {\em edge servers} in the network, and $\Set{V} = \{ \Set{V}_\text{a}, \Set{V}_\text{b} \}$. The \acp{ue} can communicate with the edge cloud via wireless channels, while wired connections are constructed between nearby edge servers and the core cloud;\footnote{While, in line with most works on \ac{mec}, we do not consider cooperation between \acp{ue}, i.e., we assume \acp{ue} do NOT compute/transmit/receive packets irrelevant to their own, it is straightforward to extend the proposed model to include cooperation between \acp{ue}.}
wireless and wireline links are collected in $\Set{E}_\text{a}$ and $\Set{E}_\text{b}$, respectively. A communication link with node $i$ as the transmitter and $j$ as the receiver is denoted by $(i,j) \in \Set{E} \triangleq \{\Set{E}_\text{a}, \Set{E}_\text{b}\}$. The incoming and outgoing neighbors of node $i$ are collected in the sets $\delta^-_i$ and $\delta^+_i$, respectively; specially, $\tilde{\delta}^+_i$ denotes the set of {\em wireless} outgoing neighbors of node $i$.

Time is divided into slots of appropriate length $\tau$, chosen such that
the uncontrollable processes (\ac{csi}, packet arrivals, etc) are \ac{iid} across time slots.
Each time slot is divided into three phases. In the {\em sensing phase}, neighboring nodes exchange local information (e.g., queue backlog) and collect \ac{csi}. Then, in the {\em outgoing phase}, decisions on task offloading, resource allocation, and packet scheduling are made and executed by each node.
Finally, during the {\em incoming phase}, each node receives incoming packets from neighbor nodes, local processing unit, and (possibly) sensing equipment.

The following parameters characterize the available computation
resources in the \ac{mec} network:
\begin{itemize}
  \item $\Set{K}_i = \{1,\cdots,K_i\}$: the possible levels of computational resource that can be allocated at node $i$;
  \item $C_{k_i}$: the compute capability (e.g., computing cycles) when $k_i\in \Set{K}_i$ resources are allocated at node $i$;
  \item $s_{k_i}$: the setup cost to allocate $k_i\in \Set{K}_i$ computational resources at node $i$;
  \item $c_{\text{pr},i}$: the computation unit operational cost (e.g., cost per computing cycle) at node $i$.
\end{itemize}
Similarly, for the {\em wireline transmission} resources on links $\Set{E}_{\text{b}}$: 
\begin{itemize}
  \item $\Set{K}_{ij} = \{1,\cdots,K_{ij}\}$: the possible levels of transmission resources that can be allocated on link $(i,j)$;
  \item $C_{k_{ij}}$: the transmission capability (e.g., bits per second) of $k_{ij} \in \Set{K}_{ij}$ transmission resources on link $(i,j)$;
  \item $s_{k_{ij}}$: the setup cost of allocating $k_{ij}\in \Set{K}_{ij}$ transmission resources on link $(i,j)$;
  \item $c_{\text{tr},ij}$: the transmission unit operational cost (e.g., cost per packet) on link $(i,j)$.
\end{itemize}

We denote by
$\V{k}(t) = \{\{k_i(t):i\in\Set{V}\},\,\{k_{ij}(t):(i,j)\in\Set{E}_{\text{b}}\}\}$ the resource allocation vector at time $t$.

Finally, we assume that each node $i$ has a maximum power budget $P_i$ for {\em wireless transmission}, and each unit of energy consumption leads to a cost of $c_{\text{wt},i}$ at node $i$. More details on the wireless transmission are presented in the next subsection.

\subsection{Wireless Transmission Model}

This section focuses on the wireless transmissions between the \acp{ue} and edge cloud. 
We employ the channel model proposed in \cite{AdhNamAhnCai:J13}, as is depicted in Fig. \ref{fig:mec}. Massive antennas are deployed at each edge server, and with the aid of beamforming techniques, it can transmit/receive data to/from multiple \acp{ue} simultaneously using the same band. The interference between different links can be neglected, given that the \acp{ue} are spatially well separated. Different edge servers are separated by a frequency division scheme, where a bandwidth of $B$ is allocated to each edge server. Each \ac{ue} is assumed to associate with only one edge server at a time.

The transmission power $p_{ij}(t)$ of each wireless link $(i,j)$ is assumed to be constant during a time slot, and we define $\V{p}_i(t) = \{p_{ij}(t):j\in\tilde{\delta}_i^+\}$ as the power allocation decision of node $i$. The packet rate of link $(i,j)$ then follows
\begin{align}\label{eq:data_rate}
R_{ij}(t) = (B/F) \log_2 \big( 1 + g_{ij}(t) p_{ij}(t) \big/ \sigma_{ij}^2 \big) , 
\end{align}
where $g_{ij}(t)$ denotes the channel gain, which is assumed to be \ac{iid} over time; $\sigma_{ij}^2$ is the noise power.
We denote by $R_{ij}(\V{p}_i, \V{g}_i)$ the data rate on link $(i, j)$ given that node $i$ adopts the transmission vector $\V{p}_i$ under the \ac{csi} $\V{g}_i = \{ g_{ij}: j\in \tilde{\delta}_i \}$.
Recall that each node $i$ has a maximum transmission power of $P_i$, and it follows
\begin{align}\label{eq:max_power}
\sum\nolimits_{j\in \tilde{\delta}_i^+}{ p_{ij}(t) } \leq P_i,\quad \forall i\in \Set{V}.
\end{align}
For each \ac{ue} $i$, a binary vector $\V{x}_i(t) = \{ x_{ij}(t):j\in\tilde{\delta}_i^+ \}$ is defined to indicate its decision of task offloading, where the link $(i,j)$ is activated if the corresponding element $x_{ij}(t) = 1$. It then follows that
\begin{align}\label{eq:association}
\sum\nolimits_{j\in \tilde{\delta}_i^+}{x_{ij}(t)} \leq 1,\quad \forall i\in \Set{V}_\text{a}.
\end{align}
Finally, we define the aggregated vectors as $\V{x}(t) = \{\V{x}_i(t):i\in\Set{V}_\text{a}\}$ and $\V{p}(t) = \{\V{p}_i(t):i\in\Set{V}\}$, respectively.

\begin{remark}\label{remark:quasi_stationary}
The length of each time slot $\tau$ is set to match the channel coherence time, which is on the order of milliseconds for millimeter wave communication (e.g., slowly moving pedestrian using sub-$6$GHz band).
On this basis, the change in network topology (and hence path loss) is assumed to be negligible between adjacent time slots.
\end{remark}
\begin{remark}
While out of the scope of this paper,
the use of wireless transmissions via broadcast can potentially increase route diversity, thus enhancing network performance when aided by techniques such as superposition coding \cite{FenLloTulMol:J18b}.
\end{remark}

\subsection{Service Model}

Suppose that the possible services that \acp{ue} might request are collected in set $\Phi$, and each service $\phi \in \Phi$ is completed by sequentially performing $M_{\phi} - 1$ tasks on the input data stream. All data streams are assumed to be composed of packets of equal size $F$, which can be processed individually.
We denote by $a_i^{(\phi)}(t)$ the number of exogenous packets of service $\phi$ arriving at node $i \in \Set{V}_\text{a}$ at time $t$, which is assumed to be \ac{iid} across time, with mean value $\lambda_i^{(\phi)}$.

We denote by $\xi_{\phi}^{(m)}$ the scaling factor of the $m$-th function of service $\phi$, i.e., the ratio of the output stream size to the input stream size; and by $r_{\phi}^{(m)}$ its workload, i.e., the required computational resource (e.g., number of computing cycles) to process a unit of input data stream.

For a given service $\phi$, a stage $m$ packet refers to a packet that is part of the output stream of function $(m-1)$, or the input stream of function $m$. Hence, a service input packet (exogenously arriving to the network) is a stage $1$ packet, and a service output packet (fully processed by the sequence of $M_{\phi}-1$ functions) is a stage $M_{\phi}$ packet.

\subsection{Queueing System}

Based on the service model introduced in the previous subsection, the state of a packet is completely described by a $3$-tuple $(u,\phi,m)$, where $u$ is the destination of the packet (i.e., the user requesting the service), $\phi$ is the requested service, $m$ is the current service stage. Packets with the same description $(u,\phi,m)$ are considered equivalent and collected in the {\em commodity} $(u,\phi,m)$. A distinct queue is created for each commodity at every node $i$, with queue length denoted by $Q_i^{(u,\phi,m)}(t)$.
Define $\V{Q}(t) = \{ Q_i^{(u,\phi,m)}(t) \}$.

To indicate the number of packets that each node {\em plans} to compute and transmit, corresponding flow variables are defined.\footnote{The planned flow does NOT take the number of available packets into consideration, i.e., a node can plan to compute/transmit more packets than it actually has. It is defined in this way for mathematical convenience.} 
For commodity $(u,\phi,m)$, we denote by $\mu_{i,\text{pr}}^{(u,\phi,m)}(t)$ the number of packets node $i$ plans to send to its \ac{cpu}, $\mu_{\text{pr},i}^{(u,\phi,m)}(t)$ the number of packets node $i$ expects to collect from its \ac{cpu}, and $\mu_{ij}^{(u,\phi,m)}(t)$ the number of packets node $i$ plans to transmit to node $j$.
According to the definitions, all the flow variables, which are collected in $\V{\mu}(t)$, must satisfy: 1) non-negativity, $\V{\mu}(t) \geq 0$; 2) {\em service chaining} constraints,
\begin{align}\label{eq:flow_consv_pr}
\mu_{\text{pr},i}^{(u,\phi,m+1)}(t) = \xi_{\phi}^{(m)} \mu_{i,\text{pr}}^{(u,\phi,m)}(t),\quad \forall i\in\mathcal V;
\end{align}
3) capacity constraints,
\begin{subequations}\begin{align}
& \sum\nolimits_{(u,\phi,m)}{ \mu_{i,\text{pr}}^{(u,\phi,m)}(t) r_\phi^{(m)} } \leq C_{k_i(t)},\quad \forall i\in\Set{V} \label{eq:cap_1} \\
& \sum\nolimits_{(u,\phi,m)}{ \mu_{ij}^{(u,\phi,m)}(t) } \leq \begin{cases} C_{k_{ij}(t)} & \forall (i,j)\in\Set{E}_{\text{b}} \\ R_{ij}(t) \tau & \forall (i,j)\in\Set{E}_{\text{a}} \end{cases} \label{eq:cap_2}
\end{align}\end{subequations}
and 4)  boundary conditions, i.e., for $\forall i \in \Set{V}$, $u\in \Set{V}_{\text{a}}$, and $(i,j)\in \Set{E}_{\text{a}}$, we force
\begin{align}\label{eq:bound_con}
\mu_{\text{pr},i}^{(u,\phi,1)}(t) = \mu_{i,\text{pr}}^{(u,\phi,M_{\phi})}(t) = \mu_{ij}^{(i,\phi,M_{\phi})}(t) = 0.
\end{align}

The queueing dynamics is then given by\footnote{The inequality is due to the definition of {\em planned} flow, i.e., the last line in \eqref{eq:queue_dynamics} can be larger than the packets that node $i$ actually receives.}
\begin{align}\label{eq:queue_dynamics}
Q_i^{(u,\phi,m)}(t+1)
& \leq \Big[ Q_i^{(u,\phi,m)}(t) - \mu_{i,\text{pr}}^{(u,\phi,m)}(t) - \sum_{j\in\delta^+_i}{\mu_{ij}^{(u,\phi,m)}(t)} \Big]^+ \nonumber \\
& \quad + \mu_{\text{pr},i}^{(u,\phi,m)}(t) + \sum\nolimits_{j\in\delta^-_i}\mu_{ji}^{(u,\phi,m)}(t) + a_i^{(u,\phi,m)}(t)
\end{align}
where $[\,\cdot\,]^+ \triangleq \max\{\,\cdot\,,0\}$. For the exogenous packets, $a_i^{(i,\phi,1)}(t) = a_i^{(\phi)}(t)$ and $0$ otherwise (i.e., $u\ne i$ or $m\ne 1$). Specially, $Q_i^{(i,\phi,M)}(t) = 0$ for completely processed packets.

\subsection{Problem Formulation}

Two metrics will be considered to evaluate the performance of the \ac{mec} network, i.e., the {\bf operation cost} and the {\bf average delay}.
More concretely, the instantaneous operation cost of the entire network at time $t$ is given by
\begin{align}
h_1(t) & =
\sum\nolimits_{i\in\Set{V}}{ \Big[ s_{k_{i}(t)} + c_{\text{pr},i} \sum\nolimits_{(u,\phi,m)}{ r_{\phi}^{(m)} \mu_{i,\text{pr}}^{(u,\phi,m)}(t) } \Big] } \nonumber \\
& \quad + \sum\nolimits_{(i,j)\in\Set{E}_\text{b} }{ \Big[ s_{k_{ij}(t)} + c_{\text{tr},ij} \sum\nolimits_{(u,\phi,m)}{ \mu_{ij}^{(u,\phi,m)}(t) } \Big] } \nonumber \\
& \quad + \sum\nolimits_{i\in\Set{V}}{ c_{\text{wt},i} \sum\nolimits_{j\in \tilde{\delta}_i^+}{x_{ij}(t) p_{ij}(t) \tau } }, 
\end{align}
which includes the costs of computation, wired transmission, and wireless transmission.
On the other hand, the average delay\footnote{Note that various service functions can change the size of the data stream differently, and we calculate the average delay using not the output, but the input stream size as the weight, out of consideration for fairness.} $\overline{h_2}$ is derived according to Little's theorem \cite{Nee:B10} as
\begin{align}\label{eq:average_delay}
\overline{h_2} = \V{\kappa}^{\operatorname{T}} \overline{ \big\{ \V{Q}(t) \big\} }
= \sum\nolimits_{i, (u,\phi,m)}{ \kappa_i^{(u,\phi,m)} \overline{ \big\{ Q_i^{(u,\phi,m)}(t) \big\} } }
\end{align}
where
\begin{align}
\kappa_i^{(u,\phi,m)} \triangleq \frac{1}{ \prod_{z=1}^{m-1}{\xi_{\phi}^{(z)}} \sum_{i\in\Set{V}_\text{a},\, \phi\in\Phi} \lambda_i^{(\phi)} },
\end{align}
and
\begin{align}
\overline{ \{ x(t) \} } = \lim_{T\to\infty}{ \frac{1}{T} \sum_{t=1}^{T}{ \E{x(t)} } }
\end{align}
denotes the expected long-term average of the random process $\{ x(t): t\geq 1 \}$. 

The optimal \ac{mec} network control problem is then formulated as making operation decisions $\{\V{x}(t), \V{k}(t), \V{p}(t), \V{\mu}(t)\}$ over time $t = 1, 2, \cdots$, such that
\begin{subequations}\label{eq:opt_pro}
\begin{align}
& \min \quad \overline{h_1} \triangleq \overline{ \{ h_1(t) \} } \label{eq:obj_function} \\
& \operatorname{s.t.} \quad\, \overline{ h_2 } < \infty \label{eq:delay_con} \\
& \hspace{.39in} \eqref{eq:data_rate}-\eqref{eq:queue_dynamics}
\end{align}
\end{subequations}
where \eqref{eq:delay_con} is a necessary condition for the \ac{mec} network to operate properly.
Bearing this goal in mind, we propose a parameter-dependent control policy which achieves suboptimal operation cost $\overline{h_1}$, while yielding a delay performance $\overline{h_2}$ that is explicitly bounded beyond \eqref{eq:delay_con}.
In addition, we can tune the parameter to strike different trade-off between the two metrics.
Details are presented in Section \ref{sec:performance_analysis}.

\section{\ac{mec} Network Capacity Region}\label{sec:cap_region}

In this section, we present a characterization for the capacity region $\Lambda$ of the \ac{mec} network, assuming that the network topology is static (which approximates the actual network with slowly-changing topology over time).
The capacity region is defined as the set of arrival vectors $\V{\lambda} = \{ \lambda_i^{(\phi)} : i\in \Set{V}_\text{a},\, \phi\in\Phi\}$ that makes problem \eqref{eq:opt_pro} feasible, i.e., there exists some control algorithm to stabilize the network.

\begin{theorem}\label{theorem:capacity_region}
An arrival vector $\V{\lambda}$ is within the capacity region $\Lambda$ if and only if there exist flow variables $\V{f} \geq 0$,
probability values $\big\{ \alpha^{(i)}_{k_i} : k_i\in \Set{K}_i \big\}$, $\big\{ \alpha^{(ij)}_{k_{ij}} : k_{ij}\in \Set{K}_{ij} \big\}$,
conditional \ac{pdf} $\psi^{(i)}_{\V{p}_i | \V{g}_i}$,\footnote{We denote by $\V{g}_i = [g_{ij}: j\in \tilde{\delta}_i]$ the possible \acp{csi} that can be observed by node $i$.
The \ac{pdf} of the transmission vector satisfies $\psi^{(i)}_{\V{p}_i | \V{g}_i} = \psi^{(i)}_{\V{p}_i | \V{g}_i} \mathbb{I}\{\V{p}_i \in \Set{P}_i \}$ with $\Set{P}_i$ given by \eqref{eq:feasible_power}, i.e., we only allow $\V{p}_i$ to take feasible values in practice.}
and conditional probabilities\footnote{
	The sample space of the conditional probability includes {\em empty packets}, accounting for the situation where no packet is computed/transmitted. We clarify that in Theorem \ref{theorem:capacity_region}, only non-empty packets are considered, which implies that $\sum_{(u,\phi,m)}{ \ell^{(i)}_{(u,\phi,m)|k_i} },\, \ell^{(ij)}_{(u,\phi,m)|k_{ij}},\, \text{or } \ell^{(ij)}_{(u,\phi,m)|\V{p}_i, \V{g}_i} \leq 1$.
}
$\big\{ \ell^{(i)}_{(u,\phi,m)|k_i}: \forall (u,\phi,m)\big\}$, $\big\{ \ell^{(ij)}_{(u,\phi,m)|k_{ij}}: \forall (u,\phi,m)\big\}$, $\big\{ \ell^{(ij)}_{(u,\phi,m)|\V{p}_i, \V{g}_i} : \forall (u,\phi,m)\big\}$
for any $i\in \Set{V}$, $(i,j) \in \Set{E}$, $k_i \in \Set{K}_i$, $k_{ij} \in \Set{K}_{ij}$ and any commodity $(u,\phi,m)$, such that
\begin{subequations}\label{eq:cap_region}
\begin{align}
& f_{\text{pr},i}^{(u,\phi,m)} + \sum_{j\in\delta_i^-}{ f_{ji}^{(u,\phi,m)} } + \lambda_i^{(u,\phi,m)} \leq f_{i,\text{pr}}^{(u,\phi,m)} + \sum_{j\in\delta_i^+}{ f_{ij}^{(u,\phi,m)} } \label{eq:capreg_1} \\
& f_{\text{pr},i}^{(u,\phi,m+1)} = \xi_\phi^{(m)} f_{i,\text{pr}}^{(u,\phi,m)} \label{eq:capreg_2} \\
& f_{i,\text{pr}}^{(u,\phi,m)} \leq \frac{1}{r_\phi^{(m)}}\sum_{k_i\in \Set{K}_i}{ \alpha_{k_i}^{(i)} \ell_{(u,\phi,m)|k_i}^{(i)} C_{i, k_i} } \label{eq:cap_3} \\
& f_{ij}^{(u,\phi,m)} \leq \sum_{k_{ij}\in \Set{K}_{ij}}{ \alpha_{k_{ij}}^{(ij)} \ell_{(u,\phi,m)|k_{ij}}^{(ij)} C_{ij, k_{ij}} } \label{eq:cap_4} \\
& f_{ij}^{(u,\phi,m)} \leq \tau \int_{\Set{P}_i} \mathbb{E}_{\V{g}_i}\left\{ \ell_{(u,\phi,m)|\V{p}_i, \V{g}_i}^{(ij)} \psi^{(i)}_{\V{p}_i | \V{g}_i} R_{ij}( \V{p}_i, \V{g}_i ) \right\} \operatorname{d} \V{p}_i \label{eq:cap_5} \\
& f_{\text{pr},i}^{(u,\phi,1)} = f_{i,\text{pr}}^{(u,\phi,M_{\phi})} = f_{uj}^{(u,\phi,M_{\phi})} = 0
\end{align}
\end{subequations}
where \eqref{eq:cap_4} and \eqref{eq:cap_5} are for wired and wireless transmission, respectively; and $\int$ in \eqref{eq:cap_5} denotes the multiple integral with domain $\Set{P}_i$, which is defined as
\begin{align}\label{eq:feasible_power}
\Set{P}_i = \begin{cases} \Big\{ \V{p}_i\geq 0 : \sum_{j\in \tilde{\delta}_i} p_{ij} \leq P_i \Big\} & i\in \Set{V}_\text{b} \\
 \Big\{ \V{p}_i\geq 0 : \sum_{j\in \tilde{\delta}_i} p_{ij} \leq P_i, \text{ only one element of }\V{p}_i\text{ is non-zero}\Big\} & i\in \Set{V}_\text{a}
\end{cases}.
\end{align}

In addition, the stationary randomized policy $*$ defined by the above values satisfies
\begin{align}
\lim_{t\to\infty} \frac{1}{T} \sum_{t=1}^{T} h( \V{D}^*(t) ) = h^\star(\V{\lambda})
\end{align}
where the decision variable $\V{D}(t) = [\V{x}(t), \V{k}(t), \V{\mu}(t), \V{p}(t)]$ and $D^*(t)$ denotes the decisions made by $*$, and $h^\star(\V{\lambda})$ denotes the optimal cost that can be achieved under the arrival rate $\V{\lambda}$ while stabilizing the network.
\end{theorem}
\begin{IEEEproof}
See Appendix \ref{apdx:cap_proof}.
\end{IEEEproof}

The stationary randomized policy mentioned in above theorem makes decisions only depending on the current observation of the uncontrollable process.
In every time slot, the policy select the {\em resource assignment decision} randomly according to \ac{pdf} $\big\{ \alpha_{k_i}^{(i)} : k_i \in \Set{K}_i \big\}$ (processing), $\big\{ \alpha_{k_{ij}}^{(ij)} : k_{ij} \in \Set{K}_{ij} \big\}$ (wired transmission) and $\psi^{(i)}_{\V{p}_i | \V{g}_i}$ (wireless transmission, which depends on the observed \acp{csi} $\V{g}_i$);
besides, a random {\em commodity} $(u,\phi,m)$ is selected by $\ell^{(i)}_{(u,\phi,m)|k_i}$, $\ell^{(ij)}_{(u,\phi,m)|k_{ij}}$ and $\ell^{(ij)}_{(u,\phi,m)|\V{p}_i}$ to perform the corresponding operation (processing or transmission).

\section{MEC Network Control}

An equivalent expression of the above problem is to replace \eqref{eq:delay_con} by (noting \eqref{eq:average_delay} and the boundedness of $\V{\kappa}$)
\begin{align}\label{eq:stability}
\overline{ \big\{Q_i^{(u,\phi,m)}(t)\big\} } < \infty,\quad \forall i,\,(u,\phi,m).
\end{align}
which is within the scope of {\em Lyapunov optimization} theory.

\subsection{\ac{ldp}}

Let the Lyapunov function of the \ac{mec} queuing system be defined as
\begin{align}
L(t) = \frac{\V{Q}(t)^{\operatorname{T}} \operatorname{diag}\{\V{\kappa}\} \V{Q}(t)}{2}
\end{align}
where $\operatorname{diag}\{\V{\kappa}\}$ denotes a diagonal matrix with $\V{\kappa}$ as its elements.
The standard procedure of the Lyapunov optimization is to 1) observe the current queue status $\V{Q}(t)$, as well as the \ac{csi} $\V{g}(t) = \{g_{ij}(t):(i,j)\in \Set{E}_{\text{a}}\}$, and then 2) minimize an upper bound of
\begin{align}
\ac{ldp} \triangleq \Delta(t) + V h_1(t) = [ L(t+1) - L(t) ] + V h_1(t)
\end{align}
where parameter $V$ controls the tradeoff between the drift $\Delta(t)$ and penalty $h_1(t)$, and the upper bound is given by\footnote{The inequality holds under the mild assumption that there exists a constant $A_{\max}$ that bounds the arrival process, i.e., $ a_i^{(\phi)}(t) \leq A_{\max}$ for $\forall\, i, \phi, t$.}
\begin{align}\label{eq:LDP_bound_1}
\text{LDP} & \leq B_0 + \V{\lambda}^{\T} \tilde{\V{Q}}(t) - \sum_{(u,\phi,m)} \Big\{ \nonumber \\
& \quad \sum_{i\in \Set{V}}{ \big[ \big( w_i^{(u,\phi,m)} - V c_{\text{pr},i} \, r_{\phi}^{(m)} \big) \mu_{i,\text{pr}}^{(u,\phi,m)}(t) - V s_{k_i(t)} \big] }  \nonumber \\
& \quad \hspace{-0.03in} + \sum_{(i,j)\in \Set{E}_\text{b} }{ \big[ \big( w_{ij}^{(u,\phi,m)} - V c_{\text{tr},ij} \big) \mu_{ij}^{(u,\phi,m)}(t) - V s_{k_{ij}(t)} \big] } \nonumber \\
& \quad \hspace{-0.03in} + \sum_{(i,j)\in \Set{E}_\text{a} }{ \big[ w_{ij}^{(u,\phi,m)} \mu_{ij}^{(u,\phi,m)}(t) -  V c_{\text{wt},i} \, p_{ij}(t) \tau \big] } \Big\}
\end{align}
where
\begin{align}\label{eq:local_delay}
\tilde{\V{Q}}(t) = \operatorname{diag}\{\V{\kappa}\}\V{Q}(t)
\end{align}
and the weights are given by
\begin{subequations}\begin{align}
\label{eq:proc_weight} w_i^{(u,\phi,m)} & = \big[ \tilde{Q}_i^{(u,\phi,m)}(t) - \xi_{\phi}^{(m)}\tilde{Q}_i^{(u,\phi,m+1)}(t) \big]^+ \\
\label{eq:tran_weight} w_{ij}^{(u,\phi,m)} & = \big[ \tilde{Q}_i^{(u,\phi,m)}(t) - \tilde{Q}_j^{(u,\phi,m)}(t) \big]^+;
\end{align}\end{subequations}
and $B_0$ is a constant that is irrelevant to queue status and decision variables (see \cite{FenLloTulMol:J18a} for the details of derivation).

The control algorithm developed in this paper aims to minimize the upper bound \eqref{eq:LDP_bound_1} based on observed $\V{Q}(t)$ and $\V{g}(t)$, which includes three parts (last three lines), relating to computation, wired transmission, and wireless transmission, respectively.
Each part can be optimized separately, with the solutions to the first two parts provided in \cite{FenLloTulMol:J18a} and the third problem addressed in the following.

The goal is to minimize the last term in \eqref{eq:LDP_bound_1} subject to \eqref{eq:max_power}, \eqref{eq:association}, and \eqref{eq:cap_2}. Note that the objective function is linear in $\V{\mu}(t)$, which leads to {\em Max Weight} mannered solution, i.e., finding the commodity with largest weight: 
\begin{align}\label{eq:opt_weight}
(u,\phi,m)^\star = \argmax_{(u,\phi,m)}\, { w_{ij}^{(u,\phi,m)} },\
w_{ij}^{\star} = w_{ij}^{(u,\phi,m)^\star}.
\end{align}
If $w_{ij}^{\star} = 0$, no packets will be transmitted, and thus no power will be allocated; otherwise, the optimal flow assignment is
\begin{align}\label{eq:opt_flow}
\big[ \mu_{ij}^{(u,\phi,m)}(t) \big]^\star = R_{ij}(t) \tau
\end{align}
when $(u,\phi,m)=(u,\phi,m)^\star$, and $0$ otherwise. Substituting the above result into the objective function leads to a reduced problem \ac{wrt} $\V{x}(t)$ and $\V{p}(t)$, i.e.,
\begin{align}\label{eq:opt_wireless}
\min\  \sum\nolimits_{ (i,j) \in \Set{E}_\text{a} }{ x_{ij}(t) \tau \big[ Vc_{\text{wt},i} \, p_{ij}(t) - w_{ij}^\star R_{ij}(t) \big] }
\end{align}
subject to \eqref{eq:max_power} and \eqref{eq:association}. Note that we can rearrange the above objective function according to the transmitting node as
\begin{align}\label{eq:opt_wireless_2}
\tau \sum\nolimits_{i\in\Set{V}}{ \Big\{ \sum\nolimits_{j\in \tilde{\delta}_i^+}{ x_{ij}(t) \big[ Vc_{\text{wt},i} \, p_{ij}(t) - w_{ij}^\star R_{ij}(t) \big] } \Big\} }
\end{align}
which enables separate decision making at different nodes.
The optimal solution for the \acp{ue} and edge servers are presented in next section, depending on the following proposition.

\begin{proposition}\label{prop:cvx}
The solution to the following problem
\begin{align}\begin{split}
& \min_{\V{p}_i(t)}\  \sum\nolimits_{ j\in \tilde{\delta}_i^+ }{ \big[ Vc_{\text{wt},i} \, p_{ij}(t) - w_{ij}^\star R_{ij}(t) \big] },\ \operatorname{s.t.}\ \eqref{eq:max_power}
\end{split}\end{align}
is $p_{ij}^\star(t) = \big[ w_{ij}^\star B/F/(V c_{\text{wt},i} + \varrho^\star) - \sigma_{ij}^2/g_{ij}(t) \big]^+$,
where $\varrho^\star$ is the minimum positive value that makes \eqref{eq:max_power} satisfied.
\end{proposition}

\subsection{\ac{mecnc} Algorithm}

In this section, we present the \ac{mecnc} algorithm, which optimizes \eqref{eq:LDP_bound_1} in a fully distributed manner.

\subsubsection{Computing Decision}

For each node $i\in\Set{V}$:
\begin{itemize}
  \item Calculate the weight for each commodity:
  \begin{align}\label{eq:cpt_w_1}
  W_i^{(u,\phi,m)} = \big[ w_i^{(u,\phi,m)}/r_{\phi}^{(m)} - Vc_{\text{pr},i} \big]^+;
  \end{align}
  \item Find the commodity $(u,\phi,m)$ with the largest weight:
  \begin{align}
  (u,\phi,m)^\star = \argmax\nolimits_{(u,\phi,m)}\, {W_i^{(u,\phi,m)}}
  \end{align}
  \item The optimal choice for computing resource allocation is
  \begin{align}
  k_i^\star(t) = \argmax\nolimits_{k_i\in\Set{K}_i} { \big[ W_i^{(u,\phi,m)^\star} C_{k_i} - V s_{k_i} \big] }
  \end{align}
  \item The optimal flow assignment is
  \begin{align}
  \hspace{-0.1in} \big[ \mu_{i,\text{pr}}^{(u,\phi,m)}(t) \big]^\star = \big[ C_{k_i^\star(t)}/r_\phi^{(m)} \big] \, \mathbb{I}\big\{ W_i^{(u,\phi,m)^\star} > 0 \big\}
  \end{align}
  when $(u,\phi,m) = (u,\phi,m)^\star$, and $0$ otherwise, with $\mathbb{I}\{\cdot\}$ denoting the indicator function.
\end{itemize}

\subsubsection{Wired Transmission Decision}

For each link $(i,j) \in \Set{E}_{\text{b}}$:
\begin{itemize}
  \item Calculate the weight for each commodity:
  \begin{align}\label{eq:trs_w_1}
  W_{ij}^{(u,\phi,m)} = \big[ w_{ij}^{(u,\phi,m)} - Vc_{\text{tr},ij} \big]^+
  \end{align}
  \item Find the commodity $(u,\phi,m)$ with the largest weight:
  \begin{align}
  (u,\phi,m)^\star & = \argmax\nolimits_{(u,\phi,m)}\, { W_{ij}^{(u,\phi,m)} }
  \end{align}
  \item The optimal choice for computing resource allocation is
  \begin{align}
  \hspace{-2mm}k_{ij}^\star(t) = \argmax\nolimits_{k_{ij}\in\Set{K}_{ij}} { \big[ W_{ij}^{(u,\phi,m)^\star} C_{k_{ij}} - V s_{k_{ij}} \big] }
  \end{align}
  \item The optimal flow assignment is
  \begin{align}
  \big[ \mu_{ij}^{(u,\phi,m)}(t) \big]^\star = C_{k_{ij}^\star(t)} \, \mathbb{I}\big\{ W_{ij}^{(u,\phi,m)^\star} > 0 \big\}
  \end{align}
  when $(u,\phi,m) = (u,\phi,m)^\star$, and $0$ otherwise.
\end{itemize}

\subsubsection{Wireless Transmission Decision}

For each wireless link $(i,j) \in \Set{E}_{\text{a}}$, find the optimal commodity $(u,\phi,m)^\star$ and the corresponding weight $w_{ij}^\star$ by \eqref{eq:opt_weight}; then,
\begin{itemize}
  \item for each edge server $i\in\Set{V}_\text{b}$: determine the transmission power by Proposition \ref{prop:cvx};
  \item for each \ac{ue} $i\in\Set{V}_\text{a}$: since it only associates with one edge server and noting the fact that generally each user can only access a limited number of edge servers, we can decide the optimal edge server by brute-forced search. More concretely, for any edge server $k\in \tilde{\delta}_i^+$, assume that \ac{ue} $i$ associates with it, i.e., let $x_{ij}(t) = 1$ if $j = k$, and $0$ otherwise in \eqref{eq:opt_wireless_2}, and solve the one-variable sub-problem by Proposition \ref{prop:cvx}, which gives the optimal transmission power $p_{ik}^\star(t)$ and the corresponding objective value $\rho_{ik}^\star$. By comparing the values, the optimal edge server to associate with is $k^\star = \argmin_{k\in\tilde{\delta}_i^+} \rho_{ik}^\star$.
\end{itemize}

\subsection{Performance Analysis}\label{sec:performance_analysis}

As is introduced in Section \ref{sec:cap_region}, for a given arrival vector $\V{\lambda}$ within the capacity region $\Lambda$, the queueing system can be stabilized by some control algorithms, while achieving the optimal cost $h_1^\star(\V{\lambda})$.
It serves as a benchmark to evaluate the performance of the developed \ac{mecnc} algorithm, which is described in the following theorem.

\begin{theorem}\label{theorem:tradeoff}
For any arrival vector $\V{\lambda}$ lying in the interior of $\Lambda$, the \ac{mecnc} algorithm can stabilize the queueing system, with the average cost and delay satisfying
\begin{align}
\overline{h_1} & \leq h_1^\star(\V{\lambda}) + \frac{B_0}{V} \\
\overline{h_2} & \leq \frac{B_0}{\epsilon} + \frac{ \big[ h_1^\star(\V{\lambda} + \epsilon \V{1}) - h_1^\star(\V{\lambda}) \big] V }{\epsilon}
\end{align}
where $\epsilon \V{1}$ denotes a vector with all elements equal to $\epsilon > 0$ that satisfies $\V{\lambda} + \epsilon \V{1} \in \Lambda$ (since $\V{\lambda}$ is in the interior of $\Lambda$).
\end{theorem}

\begin{IEEEproof}
See Appendix \ref{sec:apdx2}.
\end{IEEEproof}

Since the queueing system is stabilized, the proposed algorithm is throughput optimal.
In addition, there exists an $[O(V),O(1/V)]$ tradeoff between the upper bounds of the average delay and the incurred cost. By increasing the value of $V$, the average cost can be reduced, while increasing the average delay (but still guaranteed), which accords with the motivation behind the definition of the \ac{ldp} function \eqref{eq:LDP_bound_1}.

\section{Numerical Results}

\begin{figure}[t]
\centering
\includegraphics[width=.5\textwidth]{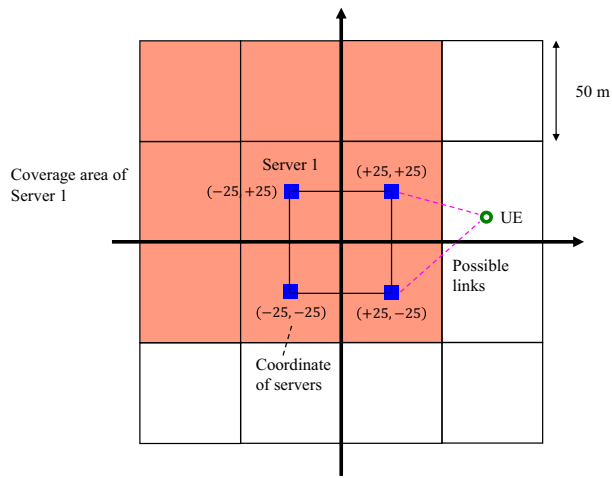}
\caption{Considered microcell with $N_\text{b} = 4$ edge servers (and $N_\text{a} = 100$ UEs).}
\label{fig:area}

\hfill
\end{figure}

Consider the square micro-cell as is shown in Fig. \ref{fig:area}, which includes $N_\text{a} = 100$ \acp{ue} and $N_\text{b} = 4$ edge servers.
Each edge server covers the \acp{ue} within the surrounding $3\times 3$ clusters.
We set the length of each time slot as $\tau = 1\ \text{ms}$.
The movements of the \acp{ue} is modeled by \ac{iid} random walk (reflecting when hitting the boundary), with the one-slot displacement distributing in Gaussian $N(0, 10^{-2} \M{I}_2)$ (the average speed of the \acp{ue} $\approx 3.9\ \text{m}/\text{s}$ under this setting).

Each user can request two services (there are two functions in each service), with the following parameter
\begin{align}
\text{Service }1 &:\ \xi_1^{(1)} = 1,\ \xi_1^{(2)} = 2;\ 1/r_1^{(1)} = 300,\ 1/r_1^{(2)} = 400 \nonumber \\
\text{Service }2 &:\ \xi_2^{(1)} = \frac{1}{3},\ \xi_2^{(2)} = \frac{1}{2};\ 1/r_2^{(1)} = 200,\ 1/r_2^{(2)} = 100 \nonumber
\end{align}
where $1/r_{\phi}^{(m)}\ [\text{Mb}/\text{CPU}]$ is the supportable input size (in one slot) given $1$ CPU resource.
The size of each packet is $F = 1\ \text{kb}$, and the number of packets arriving at each \ac{ue} is modeled by \ac{iid} Poisson processes of parameter $\lambda$.

\begin{table}
  \centering
  \caption{Available Resources and Costs of the \ac{mec} network (on the basis of second)}
  \renewcommand\arraystretch{1.5}
  \begin{tabular}{|l|c|}
    \hline
    & \ac{ue} $i\in\Set{V}_\text{a}$ \\ \hline
    Computation & $\Set{K}_i = \{0,1\},\, C_{k_i} = k_i\,\text{\acp{cpu}},\, s_{k_i} = 5k_i,\, c_{\text{pr},i} = 1\,/\ac{cpu}$  \\ \hline
    Wired Links & No wired transmission between \acp{ue} \\ \hline
    Wireless Links & $P_{i} = 200\,\text{mW},\, c_{\text{wt},i} = 1\,/\text{W}$ \\ \hline \hline
    & Edge Server $i\in\Set{V}_\text{b}$ \\ \hline
    Computation & $\Set{K}_i = \{0,\cdots, 10\},\, C_{k_i} = 5k_i\,\text{\acp{cpu}},\, s_{k_i} = 5k_i,\, c_{\text{pr},i} = .2\,/\text{CPU}$ \\ \hline
    Wired Links & $\Set{K}_{ij} = \{0, \cdots, 5\},\, C_{k_{ij}} = 10k_{ij}\,\text{Gbps},\, s_{k_{ij}} = k_{ij},\, c_{\text{tr},ij} = 1\,/\text{Gb}$  \\ \hline
    Wireless Links & $P_{i} = 10\,\text{W},\, c_{\text{wt},i} = .2\,/\text{W}$ \\ \hline
  \end{tabular}
  \label{tab:resource}
\end{table}

The available resources and corresponding costs are summarized in Table \ref{tab:resource}. For wireless transmission, millimeter wave communication is employed, operating in the band of $f_\text{c} = 30\ \text{GHz}$, and a bandwidth of $B = 100\ \text{MHz}$ is allocated to each edge server; the 3GPP path-loss model $32.4 + 20 \log_{10}(f_{\text{c}}) + 31.9 \log_{10}(\text{distance})\ \text{dB}$ for urban microcell is adopted, and the standard deviation of the shadow fading is $\sigma_{\text{SF}} = 8.2\ \text{dB}$; the antenna gain is $10\ \text{dB}$. The noise has a power spectrum density of $N_0 = -174\ \text{dBm}/\text{Hz}$.

\begin{figure}[t]
\centering
\includegraphics[width=.5\textwidth]{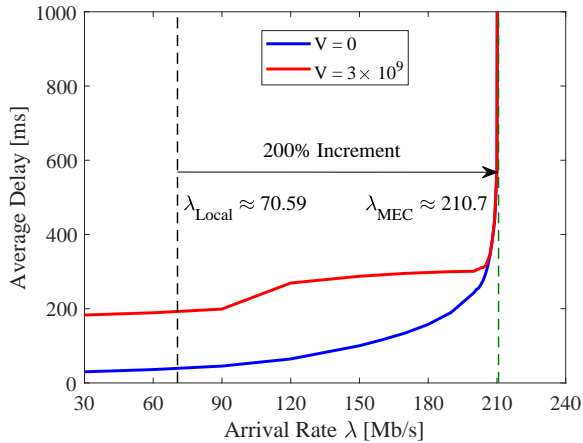}
\caption{Capacity region achieved by using $V = 0$ and $V = 3\times 10^9$.}
\label{fig:cap_region}
\end{figure}

\subsection{Capacity Region}

First, we simulate the capacity region for the described \ac{mec} network, using different values of $V$ in the \ac{mecnc} algorithm. As $\lambda$ varies, the {\em stable} average delay $\overline{\{ \V{\kappa}^\T \V{Q}(t) \} }$ is recorded (if exists) based on a long-term ($1\times 10^6$ time slots) observation of the queueing system. If the average delay is constantly growing even at the end of the time window, the average delay is defined as $\infty$, which implies that the network is not stable under the given arrival rate.

As depicted in Fig. \ref{fig:cap_region}, the average delay rises as the arrival rate $\lambda$ increases, which blows up when approaching $\lambda_{\text{MEC}} \approx 210\ \text{Mb}/\text{s}$.
This critical point can be interpreted as the boundary of the capacity region of the \ac{mec} network.
On the other hand, if all the computations are constrained to be executed at the \acp{ue}, the capacity region is reduced to $\lambda_{\text{Local}} \approx 70\ \text{Mb}/\text{s}$.\footnote{At the boundary of the capacity region, the computing capability constraint is active, i.e. $\lambda_{\text{Local}} \sum_{\phi=1}^{2}{ \sum_{m=1}^{2}{ \Xi_\phi^{(m)} w_\phi^{(m)} } } = C_{K_i}\ (i\in \Set{V}_\text{a})$.}
That is, a gain of $200\%$ is achieved with the aid of edge servers. Last but not least, note that different $V$ values lead to identical critical points, although they result in different average delay performance, which validates the throughput-optimality of the \ac{mecnc} algorithm.

\subsection{Cost-Delay Tradeoff}

\begin{figure}[t]
\centering
\begin{minipage}[t]{0.49\textwidth}
\centering
\includegraphics[width=\textwidth]{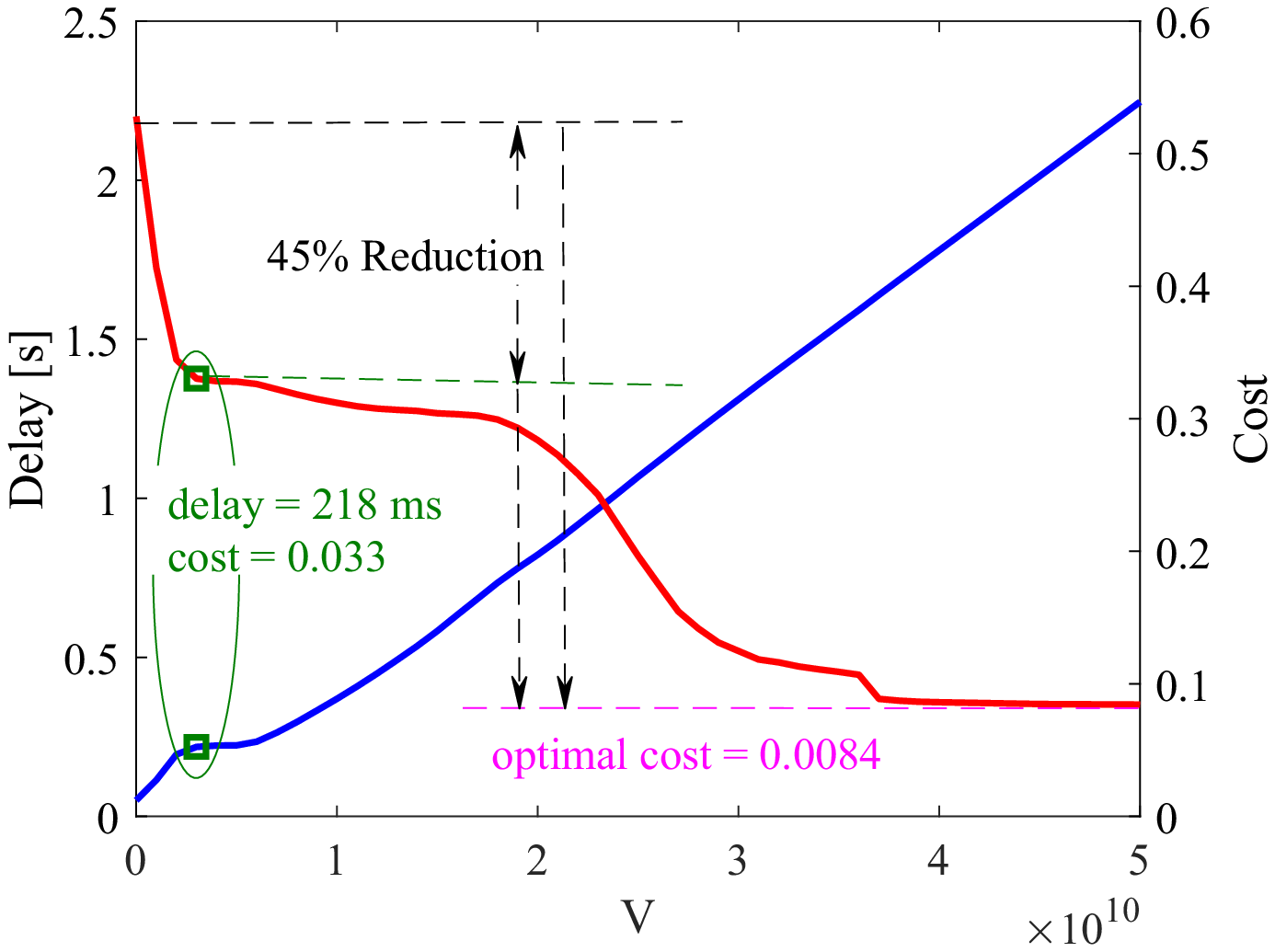}
\caption{Delay and cost performance under various $V$.}
\label{fig:tradeoff}
\end{minipage}
\hfill
\begin{minipage}[t]{0.49\textwidth}
\centering
\includegraphics[width=\textwidth]{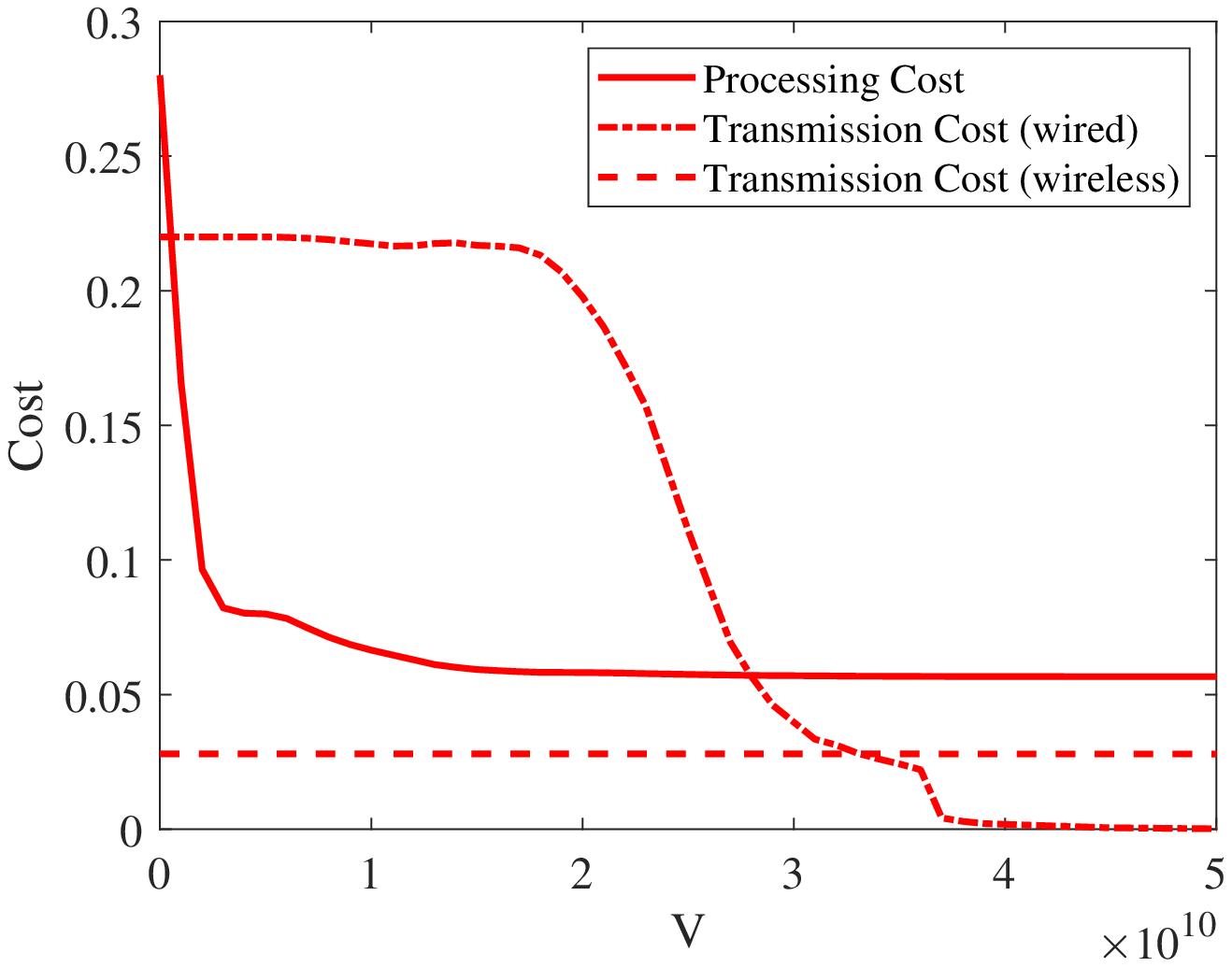}
\caption{Cost breakdown for processing and transmission part.}
\label{fig:break_down}
\end{minipage}
\end{figure}

Next, we study the delay and cost performance of the \ac{mec} network, when tuning the parameter $V$. The arrival rate is set as $\lambda = 100\ \text{Mb}/\text{s}$ (i.e., $100$ packets per slot).

The results are shown in Fig. \ref{fig:tradeoff}.
Evidently, the average delay\footnote{The result here is obtained by counting the age of the packets.} grows almost linearly with $V$, while the cost reduces as $V$ grows (with a vanishing rate at the end), which support the $[O(V),O(1/V)]$ tradeoff between the delay and cost {\em bound}.
In addition, we observe two regions of significant cost reduction, i.e., $V\in [0, 3\times 10^9]$ and $[2\times 10^{10}, 3\times 10^{10}]$.
The first reduction is due to the task offloading to the edge cloud;
while the second one results from {\em cutting off the connections between the edge servers}, which lessens the transmission cost within the edge cloud, while dramatically increasing the queueing delay (since we are not able to balance the load between the edge servers in favor of delay performance).
A more detailed cost breakdown for the cost of each part is shown in Fig. \ref{fig:break_down}.

Based on the tradeoff relationship, we can tune the value of $V$ to optimize the performance of practical \ac{mec} networks.
For example, the value $V^\star = 3 \times 10^9$ leads to an average delay of $218\ \text{ms}$, which is acceptable for real-time applications; while a cost of $0.33$, which reduces the gap to the optimal cost $0.084$ by $45\%$ (compared with $0.53$ when $V = 0$).

\begin{table}
  \centering
  \caption{Offloading Ratio under Different Values of $V$}
  \vspace{-0.1in}
  \renewcommand\arraystretch{1.5}
  \begin{tabular}{|l|c|c|c|c|}
    \hline
     & \multicolumn{2}{|c|}{Service $1$} & \multicolumn{2}{|c|}{Service $2$} \\ \hline 
    &  Function $1$ & Function $2$ & Function $1$ & Function $2$ \\ \hline 
    $V = 0$ & $20.0\%$ & $19.3\%$ & $56.7\%$ & $63.9\%$   \\ \hline
    $V = 1\times 10^9$ & $59.8\%$ & $56.3\%$ & $75.1\%$ & $91.4\%$  \\ \hline
    $V = 3\times 10^9$ & $98.8\%$ & $98.1\%$ & $99.7\%$ & $100\%$ \\ \hline
  \end{tabular}
  \vspace{-0.1in}
  \label{tab:ratio}
\end{table}

Finally, we observe the offloading ratios for different computation tasks and various $V$ values in Table \ref{tab:ratio}.
As we expect, a growing value of $V$ puts more attention to the induced cost, motivating the \acp{ue} to offload the tasks to the cloud.
In addition, we find that for all listed values of $V$, the functions of Service $2$ tend to have a higher offloading ratio.
An intuitive explanation is that Service $2$ is more compute-intensive, while resulting in lower communication overhead than Service $1$, and thus more preferable for offloading.

\section{Conclusions}

In this paper, we leveraged recent advances in the use of Lyapunov optimization theory to study the stability of computing networks, in order to address key open problems in \ac{mec} network control. 
We designed a distributed algorithm, \ac{mecnc}, that makes joint decisions about task offloading, packet scheduling, and computation/communication resource allocation.  Numerical experiments were carried out on the capacity region, cost-delay tradeoff, and task assignment performance of the proposed solution, proving to be a promising paradigm to manage next generation \ac{mec} networks.

\appendices

\section{Proof of Theorem \ref{theorem:capacity_region}}\label{apdx:cap_proof}

{\em Necessity}: We start with the discrete case, where we collect the possible \acp{csi} that can be observed by node $i$ in $\Set{G}_i$, and the transmission actions that node $i$ can take in $\Set{Z}_i$, and we assume that the cardinalities of the sets are $G$ and $Z$, respectively.
We clarify that 1) each element in both sets is a vector composed of all the information of node $i$'s outgoing links, which will be represented by its index in the set, i.e., $1\leq g\leq G$ and $1\leq z\leq Z$; 2) given the \ac{csi} and the transmission vector, the data rate of all the outgoing links are fixed, denoted by $R_{ij}(g, z)$.
The following discussions are straightforward to extend to the continuous scenario, where we push $G\to\infty$ and $Z\to\infty$ assuming uniform discretization.

Suppose there exists some control algorithm to stabilize the queueing system.
We define the following quantities within the first $t$ time slots:
\begin{itemize}
	\item $X_{i\text{p}, k}^{(u,\phi,m)}(t)$: the number of $(u,\phi, m)$-packets that are processed at node $i$ when the allocation choice is $k \in \Set{K}_{i}$, and later on successfully delivered to the destination;
	\item $X_{\text{p}i, k}^{(u,\phi,m)}(t)$: the number of $(u,\phi, m)$-packets that are received by node $i$ from its local processor when the allocation choice is $k \in \Set{K}_{i}$, and later on successfully delivered to the destination;
	\item $X_{ij, k}^{(u,\phi,m)}(t)$: the number of $(u,\phi, m)$-packets that are transmitted through wired link $(i,j)$ when the allocation choice is $k \in \Set{L}_{ij}$, and later on successfully delivered  to the destination.
	\item $X_{ij, (g, z)}^{(u,\phi,m)}(t)$: the number of $(u,\phi, m)$-packets that are transmitted through wireless link $(i,j)$ when the observed \ac{csi} is $g\in \Set{G}_i$ and the adopted transmission vector is $z\in \Set{Z}_i$, and later on successfully delivered  to the destination.
	\item $A_{i}^{(u,\phi,m)}(t)$: the number of exogenous $(u,\phi, m)$-packets that arrives at node $i$, and later on successfully delivered  to the destination.
\end{itemize}

Define the consumed computing resource
\begin{align}
S_{i, k}(t) = \sum_{(u,\phi,m)}{ r_\phi^{(m)} X_{i\text{p}, k}^{(u,\phi,m)}(t) }\quad (\forall\, k\in \Set{K}_i),\ 
S_i(t) = \sum_{k\in \Set{K}_i}{ S_{i, k}(t) }
\end{align}
for wired transmission $(i,j)\in \Set{E}_\text{b}$,
\begin{align}
X_{ij, k}(t) = \sum_{(u,\phi,m)}{ X_{ij,k}^{(u,\phi,m)}(t) }\quad (\forall\, k\in \Set{K}_{ij}),\ 
X_{ij}^{(u,\phi,m)}(t) = \sum_{k\in \Set{K}_{ij}}{ X_{ij,k}^{(u,\phi,m)}(t) }
\end{align}
and for wireless transmission $(i,j)\in \Set{E}_\text{a}$,
\begin{align}
X_{ij, (g, z)}(t) = \sum_{(u,\phi,m)}{ X_{ij, (g, z)}^{(u,\phi,m)}(t) }\quad (\forall\, g\in \Set{G}_i, z\in \Set{Z}_i),\ 
X_{ij}^{(u,\phi,m)}(t) = \sum_{(g, z) \in \Set{G}_i\times \Set{Z}_i}{ X_{ij, (g, z)}^{(u,\phi,m)}(t) }
\end{align}
In addition, we define $N_k^{(i)}(t)$ as the number of time slots when node $i$ assigns $k \in \Set{K}_i$ processing resource;
similarly, $N_k^{(ij)}(t)$ can be defined for wired transmission with $k\in \Set{K}_{ij}$.
For wireless transmission, we define $N_{(g, z)}^{(i)}(t)$ the number of time slots that the observed \acp{csi} is $g \in \Set{G}_i$, and the transmission vector $z \in \Set{G}_i$; and define $N_{g}^{(i)}(t) = \sum_{z\in \Set{Z}_i} N_{(g, z)}^{(i)}(t)$.

The following relationship then follows
\begin{subequations}\label{eq:cap_pkt}\begin{align}
& \sum_{j\in \delta_i^{-}} X_{ji}^{(u,\phi, m)}(t) + X_{\text{p}i}^{(u,\phi, m)}(t) + A_{i}^{(u,\phi, m)}(t)
= \sum_{j\in \delta_i^{+}} X_{ij}^{(u,\phi, m)}(t) + X_{i\text{p}}^{(u,\phi, m)}(t) \\
& X_{\text{p}i}^{(u,\phi, m+1)}(t)
= \xi_\phi^{(m)} X_{i\text{p}}^{(u,\phi, m)}(t) \\
& r_\phi^{(m)} X_{i\text{p}}^{(u,\phi,m)}(t)
\leq \sum_{k\in \Set{K}_i} N^{(i)}_k(t) \left( \frac{ r_\phi^{(m)} X_{i\text{p}, k}^{(u,\phi,m)}(t)}{ S_{i,k}(t) } \right) C_{i,k} \label{eq:cap_pkt_3} \\
& X_{ij}^{(u,\phi,m)}(t)
\leq \sum_{k\in \Set{K}_{ij}} N_k^{(ij)}(t) \left( \frac{X_{ij, k}^{(u,\phi,m)}(t)}{X_{ij, k}(t)} \right) C_{ij, k} \label{eq:cap_pkt_4} \\
& X_{ij}^{(u, \phi, m)}(t)
\leq \tau \sum_{(g, z) \in \Set{G}_i\times \Set{Z}_i} \left( \frac{X_{ij, (g, z)}^{(u, \phi, m)}(t)}{X_{ij, (g, z)}(t)} \right) \left( \frac{ N^{(i)}_{(g, z)}(t) }{ N^{(i)}_g(t) } \right) N^{(i)}_g(t) R_{ij}(z, g) \label{eq:cap_pkt_5}
\end{align}\end{subequations}
where \eqref{eq:cap_pkt_4} is for $(i,j) \in \Set{E}_\text{b}$ and \eqref{eq:cap_pkt_5} is for $(i,j) \in \Set{E}_\text{a}$.
The last three inequalities are interpreted as follows, which are derived from the resource constraints.
For the processing resource at node $i$, consider the subset of time slots where $k\in \Set{K}_i$ processing resource is assigned, and we  obtain
\begin{align}
S_{i,k}(t) \leq N_k^{(i)}(t) C_{ij, k}\quad (\forall\, k\in \Set{K}_i).
\end{align}
Multiply by $\big( r_\phi^{(m)} X_{i\text{p}}^{(u,\phi,m)}(t) \big) / S_{i,k}(t)$, then sum over $k\in \Set{K}_i$, and it leads to \eqref{eq:cap_pkt_3}.
Similar technique can be used to derive \eqref{eq:cap_pkt_4} and \eqref{eq:cap_pkt_5}.

Divide \eqref{eq:cap_pkt} by $t$, and push $t\to \infty$. For simplicity, assume that all the limits exist, and define
\begin{align}
& \lim_{t\to\infty} \frac{X_{i\text{p}}^{(u,\phi, m)}(t)}{t}  = f_{i\text{p}}^{(u,\phi, m)},\ 
\lim_{t\to\infty} \frac{X_{ij}^{(u,\phi, m)}(t)}{t}  = f_{ij}^{(u,\phi, m)},\ 
\lim_{t\to\infty} \frac{A_{i}^{(u,\phi, m)}(t)}{t} = \lambda_{i}^{(u,\phi, m)} \nonumber \\
& \lim_{t\to\infty} \frac{ r_\phi^{(m)} X_{i \text{p}, k}^{(u,\phi,m)}(t) }{S_{i,k}(t)}  = \ell^{(i)}_{(u,\phi, m)|k},\ 
\lim_{t\to\infty} \frac{ X_{ij, k}^{(u,\phi,m)}(t) }{X_{ij, k}(t)}  = \ell^{(ij)}_{(u,\phi, m)|k},\ 
\lim_{t\to\infty} \frac{ X_{ij, (g, z)}^{(u,\phi,m)}(t) }{X_{ij, (g, z)}(t)}  = \ell^{(ij)}_{(u,\phi, m)|(g, z)} \nonumber \\
& \lim_{t\to\infty} \frac{N^{(i)}_k(t)}{t} = \alpha^{(i)}_k,\ 
\lim_{t\to\infty} \frac{N^{(ij)}_k(t)}{t} = \alpha^{(ij)}_k,\ 
\lim_{t\to\infty} \frac{ N^{(i)}_{(g, z)}(t) }{ N^{(i)}_g(t) } = \varphi_{z|g}^{(i)},\ 
\lim_{t\to\infty} \frac{N^{(i)}_g(t)}{t} = \mathbb{P}^{(i)}_g
\end{align}
where $\mathbb{P}^{(i)}_g$ is the probability that the observed \acp{csi} by node $i$ is $g$.
Substitute the above definitions into the previous relationships, and the proof is concluded.

Furthermore, let $G\to\infty$ and $Z\to\infty$, and we assume $\varphi_{z|g}^{(i)} \to \psi_{p|g}^{(i)}$, and $\mathbb{P}^{(i)}_g$ converges to the true \acp{csi} distribution.
In this case, \eqref{eq:cap_pkt_5} becomes
\begin{align}\begin{split}
f_{ij}^{(u, \phi, m)}(t)
& \leq \tau \sum_{(g, z) \in \Set{G}_i\times \Set{Z}_i} \ell_{(u, \phi, m) | (g, z)}^{(ij)} \varphi_{z|g}^{(i)} \mathbb{P}^{(i)}_g R_{ij}(z, g) \\
& \to \tau \int_{z\in \Set{Z}_i} \ell_{(u, \phi, m) | (g, z)}^{(ij)} \psi_{z|g}^{(i)} p^{(i)}_g R_{ij}(z, g) dz dg \\
& = \tau \int_{\Set{P}_i} \mathbb{E}_{\V{g}_i} \left\{ \ell_{(u, \phi, m) | (\V{g}_i, \V{p}_i)}^{(ij)} \psi_{\V{p}_i|\V{g}_i}^{(i)} R_{ij}(\V{p}_i, \V{g}_i) \right\} d\V{p}_i
\end{split}\end{align}
where we recall that each index $z$ (or $g$) corresponds to an action (or \ac{csi}) vector.

Finally, assume that the control algorithm considered above leads to the optimal cost $h^\star(\V{\lambda})$.
By definition, we can obtain
\begin{align}
h^\star(\V{\lambda}) = \frac{1}{T} \sum_{t=1}^{T}{ h(\V{D}(t)) }
= \frac{1}{T} \sum_{\V{D}\in \Set{D}}{ N_{\V{D}}(T) h(\V{D}) }
\end{align}
where $\Set{D}$ denotes the action space, and $N_{\V{D}}(T)$ is the number of time slots that action $\V{D}$ is selected within the observed interval.
Further note that the constructed randomized policy $*$ (as is introduced in Section \ref{sec:cap_region}) satisfies that (by law of large number)
\begin{align}
\lim_{T\to\infty} \frac{N^*_{\V{D}}(T)}{T}
= \lim_{T\to\infty} \frac{N_{\V{D}}(T)}{T}
\end{align}
where $N^*_{\V{D}}(T)$ is the number of time slots that policy $*$ chooses the action $\V{D}$.
It follows that
\begin{align}
\lim_{t\to\infty} \frac{1}{T} \sum_{t=1}^{T} h( \V{D}^*(t) ) 
= \lim_{t\to\infty} \frac{1}{T} \sum_{\V{D}\in \Set{D}}{ N^*_{\V{D}}(T) h(\V{D}) }
= \lim_{t\to\infty} \frac{1}{T} \sum_{\V{D}\in \Set{D}}{ N_{\V{D}}(T) h(\V{D}) }
= h^\star(\V{\lambda})
\end{align}
Therefore, the policy $*$ also achieves the optimal cost.

{\em Sufficiency}: We consider the stationary randomized policy constructed using the probability value $\alpha$, $\psi$ and $\ell$, as is explained in Section \ref{sec:cap_region}.
The resulting flow variables $\mu_{ij}^{(u,\phi,m)}(t)$ is i.i.d. over time slots, and by \eqref{eq:cap_3} to \eqref{eq:cap_5} we have
\begin{align}
\E{ \mu_{i,\text{pr}}^{(u,\phi,m)}(t) }
& =\E{ \E{ \mu_{i, \text{pr}}^{(u,\phi,m)}(t) \middle| k_i } }
= \E{ \ell^{(i)}_{(u,\phi,m)|k_i} \left( \frac{ C_{i, k_i} }{ r_\phi^{(m)} } \right) } \nonumber \\
& = \frac{1}{ r_\phi^{(m)} }\sum_{k_i\in \Set{K}_i} \alpha^{(i)}_{k_i} \ell^{(i)}_{(u,\phi,m)|k_i} C_{i, k_i}
= f_{i, \text{pr}}^{(u,\phi,m)} \\
\E{ \mu_{ij}^{(u,\phi,m)}(t) }
& =\E{ \E{ \mu_{ij}^{(u,\phi,m)}(t) \middle| k_{ij} } }
= \E{ \ell^{(ij)}_{(u,\phi,m)|k_{ij}} C_{ij, k_{ij}} } \nonumber \\
& = \sum_{k_{ij} \in \Set{K}_{ij}} \alpha^{(ij)}_k \ell^{(ij)}_{(u,\phi,m)|k_{ij}} C_{ij, k_{ij}} = f_{ij}^{(u,\phi,m)} \\
\E{ \mu_{ij}^{(u,\phi,m)}(t) }
& = \E{ \E{ \mu_{ij}^{(u,\phi,m)}(t) \middle| \V{g}_{i} } }
= \E{ \tau \int_{\Set{P}_i} \ell_{(u,\phi,m)|\V{p}_i, \V{g}_i}^{(ij)} R_{ij}( \V{p}_i, \V{g}_i ) \psi^{(i)}_{ \V{p}_{i} | \V{g}_{i} } \operatorname{d} \V{p}_i } \nonumber \\
& = \tau \int_{\Set{P}_i} \mathbb{E}_{\V{g}_i}\left\{ \ell_{(u,\phi,m)|\V{p}_i, \V{g}_i}^{(ij)} R_{ij}( \V{p}_i, \V{g}_i ) \psi^{(i)}_{\V{p}_i | \V{g}_i} \right\} \operatorname{d} \V{p}_i
= f_{ij}^{(u,\phi,m)}
\end{align}
Besides, it follows \eqref{eq:capreg_2} that
\begin{align}
\E{ \mu_{\text{pr}, i}^{(u,\phi,m)}(t) } = \E{ \xi^{(\phi,m)} \mu_{i, \text{pr}}^{(u,\phi,m-1)}(t) } = \xi^{(\phi,m)} f_{i, \text{pr}}^{(u,\phi,m-1)} = f_{\text{pr}, i}^{(u,\phi,m)}
\end{align}
Substitute the above results into \eqref{eq:capreg_1}, we obtain
\begin{align}
\E{ \sum_{j} \mu_{ji}^{(u,\phi, m)}(t) + \mu_{\text{pr}, i}^{(u,\phi, m)}(t) + a_{i}^{(u,\phi, m)}(t) } \leq 
\E{ \sum_{j} \mu_{ij}^{(u,\phi, m)}(t) + \mu_{i,\text{pr}}^{(u,\phi, m)}(t) }
\end{align}
Recall the queueing dynamics of $Q_i^{(u,\phi, m)}(t)$, where left-side and right-side in above equation denote the incoming and outgoing packets of the queue, respectively.
It indicates that $Q_i^{(u,\phi, m)}(t)$ is mean rate stable \cite{Nee:B10}.

\section{Proof for Theorem \ref{theorem:tradeoff}}\label{sec:apdx2}

Note that the LDP expression satisfies
\begin{align}\begin{split}\label{eq:key_ineq}
\Delta(t) + Vh_1( \V{D}(t) )
\leq B_0 + Vh_1( \V{D}(t) ) \hspace{3.5in} \\
\quad - \sum_{i, (d,\phi,m)} \Big[ \sum_{j\in \delta_i^+} \mu_{ij}^{(u, \phi, m)}(t) + \mu_{i, \text{pr}}^{(u, \phi, m)}(t) - \sum_{j\in \delta_i^-} \mu_{ji}^{(u, \phi, m)}(t) - \mu_{\text{pr}, i}^{(u, \phi, m)}(t) - a_{i}^{(u, \phi, m)}(t) \Big] \tilde{Q}_i^{(u, \phi, m)}(t) \\
\leq B_0 + Vh_1( \hat{\V{D}}(t) ) \hspace{3.5in} \\
\quad - \sum_{i, (d,\phi,m)} \Big[ \sum_{j\in \delta_i^+} \hat{\mu}_{ij}^{(u, \phi, m)}(t) + \hat{\mu}_{i, \text{pr}}^{(u, \phi, m)}(t) - \sum_{j\in \delta_i^-} \hat{\mu}_{ji}^{(u, \phi, m)}(t) - \hat{\mu}_{\text{pr}, i}^{(u, \phi, m)}(t) - a_{i}^{(u, \phi, m)}(t) \Big] \tilde{Q}_i^{(u, \phi, m)}(t)
\end{split}\end{align}
where $\hat{\V{D}}$ (which includes the decisions on flow assignment $\hat{\V{\mu}}$) is the decision made by the stationary randomized policy that achieves minimum cost for arrival rate $\V{\lambda} + \epsilon \V{1}$, which can be constructed according to Appendix \ref{apdx:cap_proof} (the existence of $\epsilon$ is guaranteed since $\V{\lambda}$ is in the interior of the capacity region).

Note the following facts about the decision $\hat{\V{D}}(t)$:
1) it is a stationary randomized policy, thus the decisions, specially $\hat{\V{\mu}}(t)$, are independent with $\tilde{\V{Q}}(t)$,
2) this policy is cost optimal for arrival rate $\V{\lambda} + \epsilon \V{1}$, i.e., $\E{h(\hat{\V{D}}(t))} = h^\star(\V{\lambda} + \epsilon \V{1})$, 3) the network is stabilized using this randomized algorithm under $\V{\lambda} + \epsilon \V{1}$, thus
\begin{align}\begin{split}
\E{ \sum_{j\in \delta_i^+} \hat{\mu}_{ij}^{(u, \phi, m)}(t) + \hat{\mu}_{i, \text{pr}}^{(u, \phi, m)}(t) - \sum_{j\in \delta_i^-} \hat{\mu}_{ji}^{(u, \phi, m)}(t) - \hat{\mu}_{\text{pr}, i}^{(u, \phi, m)}(t) - \hat{a}_{i}^{(u, \phi, m)}(t) } \geq 0
\end{split}\end{align}
where $\hat{a}_{i}^{(u, \phi, m)}(t)$ is an (imaginary) arriving process with mean $\lambda_{i}^{(u, \phi, m)}+\epsilon$, which indicates
\begin{align}\begin{split}
\E{ \sum_{j\in \delta_i^+} \hat{\mu}_{ij}^{(u, \phi, m)}(t) + \hat{\mu}_{i, \text{pr}}^{(u, \phi, m)}(t) - \sum_{j\in \delta_i^-} \hat{\mu}_{ji}^{(u, \phi, m)}(t) - \hat{\mu}_{\text{pr}, i}^{(u, \phi, m)}(t) } - \lambda_{i}^{(u, \phi, m)} \geq \epsilon.
\end{split}\end{align}

Then we take expectation of \eqref{eq:key_ineq}, and use the above facts
\begin{align}\begin{split}
\E{ \Delta(t) + Vh_1( \V{D}(t) ) }
\leq B_0 + V h_1^\star(\V{\lambda} + \epsilon \V{1}) - \epsilon \sum_{i, (d,\phi,m)} \E{ \tilde{Q}_i^{(u, \phi, m)}(t) }
\end{split}\end{align}
It immediately gives (by telescoping sum, see \cite{Nee:B10} for details)
\begin{align}
\lim_{T\to\infty}\frac{1}{T} \sum_{t=0}^{T-1} \E{ h_1(\V{D}(t)) }
& \leq h_1^\star(\V{\lambda} + \epsilon \V{1}) + \frac{B_0}{V} \label{eq:opt_1} \\
\lim_{T\to\infty}\frac{1}{T} \sum_{t=0}^{T-1} \sum_{i, (d,\phi,m)} \E{ \tilde{Q}_i^{(u, \phi, m)}(t) }
& \leq \frac{B_0}{\epsilon} + \left[ \frac{ h_1^\star(\V{\lambda} + \epsilon \V{1}) - h_1^\star(\V{\lambda}) }{\epsilon} \right] V  \label{eq:opt_2}.
\end{align}
Recall the definition of $\tilde{\V{Q}}(t)$ in \eqref{eq:local_delay}, as well as its relationship to the average (queueing) delay \eqref{eq:average_delay}, and \eqref{eq:opt_2} is equivalent to
\begin{align}
\overline{h_2} = \sum_{i, (d,\phi,m)} \overline{ \left\{ \tilde{Q}_i^{(u, \phi, m)}(t) \right\} }
& \leq \frac{B_0}{\epsilon} + \left[ \frac{ h_1^\star(\V{\lambda} + \epsilon \V{1}) - h_1^\star(\V{\lambda}) }{\epsilon} \right] V.
\end{align}
On the other hand, note that \eqref{eq:opt_1} holds for every $\epsilon > 0$, and therefore it holds for a sequence $\{ \epsilon_n \}_{n>0} \downarrow 0$, which implies
\begin{align}
\overline{h_1} = \overline{ \left\{ h_1(\V{D}(t)) \right\} } \leq h_1^\star(\V{\lambda}) + \frac{B_0}{V}.
\end{align}

\ifCLASSOPTIONcaptionsoff
  \newpage
\fi

\bibliographystyle{IEEEtran}
\bibliography{IEEE_abrv,AgI}

\begin{thebibliography}{10}
\providecommand{\url}[1]{#1}
\csname url@samestyle\endcsname
\providecommand{\newblock}{\relax}
\providecommand{\bibinfo}[2]{#2}
\providecommand{\BIBentrySTDinterwordspacing}{\spaceskip=0pt\relax}
\providecommand{\BIBentryALTinterwordstretchfactor}{4}
\providecommand{\BIBentryALTinterwordspacing}{\spaceskip=\fontdimen2\font plus
\BIBentryALTinterwordstretchfactor\fontdimen3\font minus
  \fontdimen4\font\relax}
\providecommand{\BIBforeignlanguage}[2]{{%
\expandafter\ifx\csname l@#1\endcsname\relax
\typeout{** WARNING: IEEEtran.bst: No hyphenation pattern has been}%
\typeout{** loaded for the language `#1'. Using the pattern for}%
\typeout{** the default language instead.}%
\else
\language=\csname l@#1\endcsname
\fi
#2}}
\providecommand{\BIBdecl}{\relax}
\BIBdecl

\bibitem{cai2020mec}
Y.~Cai, J.~Llorca, A.~M. Tulino, and A.~F. Molisch, ``\textnormal{Mobile edge
  computing network control: Tradeoff between delay and cost},'' in \emph{Proc.
  IEEE Global. Telecomm. Conf.}, Taipei, Taiwan, Dec. 2020, pp. 1--6.

\bibitem{cai2022metaverse}
------, ``\textnormal{Compute- and data-intensive networks: The key to the
  Metaverse},'' arXiv:2204.02001. [Online]. Available:
  \url{https://arxiv.org/abs/2204.02001}, Apr. 2022.

\bibitem{mach17mecsurvey}
P.~Mach and Z.~Becvar, ``\textnormal{Mobile edge computing: A survey on
  architecture and computation offloading},'' \emph{{IEEE} Commun. Surveys
  Tuts.}, vol.~19, no.~3, pp. 1628--1656, Mar. 2017.

\bibitem{DinLeeNiyWan:J13}
H.~T. Dinh, C.~Lee, D.~Niyato, and P.~Wang, ``\textnormal{A survey of mobile
  cloud computing: architecture, applications, and approaches},'' \emph{Wirel.
  Commun. Mob. Comput.}, vol.~13, no.~18, pp. 1587--1611, Dec. 2013.

\bibitem{CheHao:J18}
M.~Chen and Y.~Hao, ``\textnormal{Task offloading for mobile edge computing in
  software defined ultra-dense network},'' \emph{{IEEE} J. Sel. Areas Commun.},
  vol.~36, no.~3, pp. 587--597, Mar. 2018.

\bibitem{cai2021delay}
Y.~Cai, J.~Llorca, A.~M. Tulino, and A.~F. Molisch, ``\textnormal{Optimal cloud
  network control with strict latency constraints},'' in \emph{Proc. IEEE Int.
  Conf. Commun.}, Montreal, Canada, Jun. 2021, pp. 1--6.

\bibitem{LyuTiaNiZhaZhaLiu:J18}
X.~Lyu, H.~Tian, W.~Ni, Y.~Zhang, P.~Zhang, and R.~P. Liu,
  ``\textnormal{Energy-efficient admission of delay-sensitive tasks for mobile
  edge computing},'' \emph{{IEEE} Trans. Commun.}, vol.~66, no.~6, pp.
  2603--2616, Jun. 2018.

\bibitem{TraPom:J19}
T.~X. Tran and D.~Pompili, ``\textnormal{Joint task offloading and resource
  allocation for multi-server mobile-edge computing networks},'' \emph{{IEEE}
  Trans. Veh. Technol.}, vol.~68, no.~1, pp. 856--868, Jan. 2019.

\bibitem{BarLloTulRam:C15}
M.~Barcelo, J.~Llorca, A.~M. Tulino, and N.~Raman, ``\textnormal{The cloud
  service distribution problem in distributed cloud networks},'' in \emph{Proc.
  IEEE Int. Conf. Commun.}, London, UK, May 2015, pp. 344--350.

\bibitem{BarChoAhmBou:C15}
M.~F. Bari, S.~R. Chowdhury, R.~Ahmed, and R.~Boutaba, ``\textnormal{On
  orchestrating virtual network functions in NFV},'' in \emph{Int. Conf. on
  Network and Service Management (CNSM)}, Barcelona, Spain, Nov. 2015, pp.
  50--56.

\bibitem{BhaJaiSamErb:J16}
D.~Bhamare, R.~Jain, M.~Samaka, and A.~Erbad, ``\textnormal{A survey on service
  function chaining},'' \emph{Journal of Network and Computer Applications},
  vol.~75, no.~1, pp. 138--155, Nov. 2016.

\bibitem{cai2022multicast_arxiv}
Y.~Cai, J.~Llorca, A.~M. Tulino, and A.~F. Molisch, ``\textnormal{Decentralized
  control of distributed cloud networks with generalized network flows},''
  arXiv:2204.09030. [Online]. Available:
  \url{https://arxiv.org/abs/2204.09030}, Apr. 2022.

\bibitem{cai2021multicast}
------, ``\textnormal{Optimal multicast service chain control: Packet
  processing, routing, and duplication},'' in \emph{Proc. IEEE Int. Conf.
  Commun.}, Montreal, Canada, Jun. 2021, pp. 1--7.

\bibitem{FenLloTulMol:J18a}
H.~Feng, J.~Llorca, A.~M. Tulino, and A.~F. Molisch, ``\textnormal{Optimal
  dynamic cloud network control},'' \emph{{IEEE/ACM} Trans. Netw.}, vol.~26,
  no.~5, pp. 2118--2131, Oct. 2018.

\bibitem{FenLloTulMol:J18b}
------, ``\textnormal{Optimal control of wireless computing networks},''
  \emph{{IEEE} Trans. Wireless Commun.}, vol.~17, no.~12, pp. 8283--8298, Dec.
  2018.

\bibitem{AdhNamAhnCai:J13}
A.~Adhikary, J.~Nam, J.-Y. Ahn, and G.~Caire, ``\textnormal{Joint spatial
  division and multiplexing -- The large-scale array regime},'' \emph{{IEEE}
  Trans. Inf. Theory}, vol.~59, no.~10, pp. 6441--6463, Oct. 2013.

\bibitem{Nee:B10}
M.~J. Neely, \emph{Stochastic network optimization with application to
  communication and queueing systems}.\hskip 1em plus 0.5em minus 0.4em\relax
  San Rafael, CA, USA: Morgan {\&} Claypool, 2010.

\end{thebibliography}

\end{document}